\documentstyle[aps]{revtex}
\newcounter{fgno}
\def\fff{\advance\baselineskip
by-1pt\refstepcounter{fgno}\bigskip\noindent\ixpt{Figure~\thefgno }}
\def\epsfbox#1{\relax}
\newdimen\epsfxsize
\def\bfig{\begin{figure}\epsfxsize\hsize}
\def\efig{\end{figure}}
\def\bfigtwo{\begin{figure*}\epsfxsize\textwidth}
\def\efigtwo{\end{figure*}}

\def\changed{\relax}
\def\sec#1\par{\section*{#1}}
\def\subsec#1\par{\subsection*{#1}}
\def\bi{\begin{list}{$\bullet$}{\topsep0pt \parsep=\parskip \itemsep=0pt}}
\def\y{\item}
\def\ei{\end{list}}
\def\bpgm{$$\halign\bgroup&{\sl##}\hfil\cr}
\def\epgm{\egroup$$}
\def\[{$$} \def\]{$$}  
\def\eqbox#1{\fbox{$ \displaystyle #1 $}}
\def\(#1\){$$\eqbox{#1}$$}   
\def\mede#1{\langle\,#1\,\rangle}

\def\part{\partial}

\title{Traffic at the edge of chaos}
\author{Kai Nagel$^{a,c,d}$ \and Steen Rasmussen$^{a,b,c}$ \\
$^a$TSA-DO/SA, MS-M997 and
$^b$CNLS, MS-B258, Los Alamos National Laboratory,\\
Los Alamos, NM~87545, U.S.A.\\
$^c$Santa Fe Institute, 1660 Old Pecos Trail, Santa Fe, NM 87505, U.S.A.\\
$^d$Zentrum f\"ur Paralleles Rechnen ZPR, Universit\"at zu K\"oln,
50923 K\"oln, Germany\footnote{%
Permanent affiliation%
}\\
{\tt kai@zpr.uni-koeln.de, steen@lanl.gov}
}
\begin{document}

\maketitle

\begin{abstract}
We use a very simple description of human driving behavior to simulate
traffic.  The regime of maximum vehicle flow in a closed system shows
near-critical behavior, and as a result a sharp decrease of the
predictability of travel time.  Since Advanced Traffic Management
Systems (ATMSs) tend to drive larger parts of the transportation system
towards this regime of maximum flow, we argue that in consequence the
traffic system as a whole will be driven closer to criticality, thus
making predictions much harder.  A simulation of a simplified
transportation network supports our argument.
\end{abstract}

\sec 1.~Introduction

More and more metropolitan areas worldwide suffer from a transportation demand
which largely exceeds capacity.
In many cases, it is not possible or, even not desirable to extend capacity to
meet the demand \cite{Meyer_Miller84}.
In consequence, a consistent management of the large, distributed, man-made
transportation systems has become more and more important.
Examples of such activities include the construction of fast mass transit
systems,
the introduction of local bus lines, design of traveler informational systems
and car pooling to improve the use of current capacity, introduction of
congestion pricing,
and in the long term also guidance of the urban planning process towards an
evolution of urban areas with lower transportation needs.

Unfortunately, the man-made transportation systems are highly complex,
which makes them very difficult to manage.  Due to the complexity of the
dynamics of these systems, control decisions often lead to
counter-intuitive results.  In fact, management measures may even have
consequences opposite to their intention.  A clear example of how this
can happen is the addition of a new street in a particular road network
which leads to a {\it reduced} overall capacity~\cite{Cohen}.  The
reason for this dynamical response is an extreme example of the general
conflict between the individual traveler's optimal travel plans (Nash
Equilibrium) and the travel plans that give overall maximal throughput;
the System Optimum~\cite{Arnott}.  At the level of a metropolitan
region, the transportation dynamics is the aggregated result of
thousands or, in some cases, millions of individual trip-making
decisions for the movement of people and goods between origins and
destinations.  And every decision is based on incomplete information of
the state of the transportation system as a whole.  Since complete
global knowledge of the current (and future) state(s) of a
transportation system seems very difficult to obtain, future
informational based control strategies probably to a large extent should
be based on self-organizing local strategies.  However, that would still
not take away the tension between global and local transportation optima
which is one of the many reasons why predictability is very difficult in
such systems.

\changed
There is another source of unpredictability which may very well become
more dominating in a foreseeable future: Assume that all these
management measures and modern information technology succeed in
moving the transportation system closer towards higher efficiency.
Then we face another problem.  In road traffic systems, there is a
critical regime around maximal capacity, as we shall see, which
implies that transportation systems are very sensitive to small
perturbations in this regime.  Small perturbations will generate large
fluctuations in congestion formation and thus travel times.

This is the topic for our paper.

One method of dealing with the inherent complexities of the large
transportation systems is to represent the systems and generate their
dynamics through simulation.  The most straightforward way seems to be
a bottom-up microsimulation of the dynamics of all travelers and loads
at the level of where the transport decisions are made.  Starting with
a generation of travel demands and trip decisions, then routing, over
traffic, eventually the consequences for congestion frequencies,
travel time, air quality etc.\ are generated and can thus be analyzed.
This is the approach used by the TRANSIMS project~\cite{Alife4}, which
this work also is a part of.  Note that all the performance properties
that we may be interested in in a transportation system (in fact in
any man-made system) are emergent properties from the interacting
objects in the system.  They are nowhere explicitly represented at the
level of the interacting objects.  They are generated through
the dynamics.

The advantage of a microsimulation approach is that the system
dynamics is being generated through the simulation with all its
emergent properties without any explicit assumptions or aggregated
models for these properties.  The major disadvantages of a complete
microsimulation are extremely high computational demands on one side
and perhaps explanatory problems on the other.  The inclusion of many
details of reality may be excellent for generating a dynamics which is
close to the system under investigation, but it does not necessarily
lead to a better understanding of the basic (minimal) mechanisms that
cause the dynamics.  Therefore the TRANSIMS project also includes the
investigation of much simpler and computationally less demanding
models as the one we are going to discuss here.

In this paper, we concentrate on an extreme case of such a simplified
transportation system.  Out of the many modes of transportation (bus,
train, $\ldots$) we only include vehicular traffic, and we assume that
all vehicles as well as drivers are of the same type.  Our model
includes only single lane traffic, and the driving behavior is modeled
by only a few very basic rules.  The travelers may have individual
routing plans so that they know the sequence of links and exits they
want to use to go from their origin to their destination on a given
transportation network.  They can also re-plan depending on their
earlier experiences of travel time.  The approach is extendable to,
e.g., multi-lane traffic and/or different vehicle
types~\cite{Rickert1,Rickert2}.

We use numerical techniques from Computational Physics (cellular
automata~\cite{Wolfram,Stauffer}), and because of this similarity
together with the resulting high computational speed~\cite{NaSchleich},
we are able to use methods of analysis originating in Statistical
Physics (critical phenomena, scaling laws)~\cite{Landau_Lifshiz}.  We
obtain results which are easy to interpret in the context of everyday
experience, which is the more surprising as it is common belief that
traffic is deeply coupled to the unpredictability of human behavior and
cannot be modeled in terms of simple cellular automata rules.

Other large transportation microsimulation projects which are also
dealing with different aspects of the dynamic complexities of large
transportation systems are PARAMICS~\cite{paramics} and
TRAFF/NETSIM~\cite{netsim}.

The main part of this paper is divided into two parts.  The first one
deals with results for ``traffic in a closed loop'', i.e.\ without
ramps or junctions.  We review recent results about the connection
between jams, maximum throughput, and critical behavior; and we
present new results about the relation of these phenomena to travel
times.  In the second part of the paper, we turn to networks.  We
concentrate on a simple (minimal) example, which is nevertheless
sufficient to discuss some of the issues we believe are important,
especially our prediction that traffic systems become more variable
when pushed (by traffic management) towards higher efficiency.  Our
simulation results support this prediction.  We finish with a
conclusion.

\sec 2.~Single lane traffic in a closed loop

\subsec 2.1~Single lane cellular automata model

Our freeway traffic model has been described in detail in Ref.~\cite{NaS92}.
Therefore, we only give a short account of the essentials.

The single lane version of the model is defined on a one-dimensional
array of length $L$, representing a (single-lane) freeway.  Each site of
the array can only be in one of the following seven states: It may be
empty, or it may be occupied by one car having an integer velocity
between zero and five.  This integer number for the velocity is the
number of sites each vehicle moves during one iteration. Before the
movement, rules for velocity adaption ensure ``crash-free'' traffic.
The choice of five as maximum velocity is somewhat arbitrary, but it can
be justified by comparison between model and real world measurements,
combined with the aim for simplicity of the model.  In any case, any
value $v_{max} \ge 2$ seems to give qualitatively the same results
(i.e.\ the emergence of branching jam waves).  For every (arbitrary)
configuration of the model, one iteration consists of the following
steps, which are each performed simultaneously for all vehicles ($gap
:=$ number of unoccupied sites in front of a vehicle):

\bi

\y {\bf Acceleration of free vehicles:} Each vehicle of speed $v <
v_{max}$ with $gap \ge v+1$ accelerates
to $v+1$: $v \to v + 1$.

\y {\bf Slowing down due to other cars:} Each vehicle (speed $v$)
with $gap \le v-1$  reduces its speed to $gap$: $v \to gap$.

\y {\bf Randomization:} Each vehicle (speed $v$) reduces its speed
by one with probability $1/2$: $v \to \max[ \, v-1 , 0 \,]$ (takes
into consideration individual fluctuations).

\y {\bf Movement:} Each vehicle advances $v$ sites.

\ei
The three first steps may be called the ``velocity update''.
The randomization step condenses three different behavioral patterns
into one single computational rule: (i)~Fluctuations at free driving,
when no other car is close; (ii)~Non-deterministic acceleration;
(iii)~Overreactions when slowing down.

Already this simple model gives realistic backtraveling disturbances,
as can be seen in the top two pictures of Fig.~1.  In addition, one
obtains a realistic fundamental diagram for, e.g., the flow~$q$ versus
the density~$\rho$.  Fig.~2 gives simulation results for: (i)~short
time averages in a large system, (ii)~long time averages in a large
system, (iii)~long time averages in a small system.  These results are
obtained for a closed system with periodic boundary conditions, i.e.\
``traffic in a closed loop''.  The small system means $L = 10^2$, a
long system has $L \ge 10^4$.
\begin{figure}[b]
\fff {\em (next page)}.
Space-time plots at different resolutions of traffic at different
densities.  {\em Left column:\/} Density~$\rho = 0.07$, slightly below
the regime of maximum flow.  {\em Right column:\/} Density~$\rho = 0.1$,
slightly above the regime of maximum flow.  Resolutions are from top to
bottom 1:1, 1:4, and 1:16.  In other words, in the top row, each pixel
corresponds to one site $(x,t)$, and one can follow the movement of
individual cars from left to right.  In the bottom row, $16 \times 16$
pixel of the space-time information are averaged to one pixel of the
plot.

\end{figure}

Measurements are done at one fixed place in the system; technically,
we measure
\[
\rho = { 1 \over T } \sum_{t=t_0}^{t_0 + T-1}
\, { 1 \over v_{max} } \sum_{x=x_0}^{x_0 + v_{max}-1} \delta(x,t)
\]
and
\[
q = { 1 \over T } n_T \ .
\]
$\delta(x,t)$ is $1$ if $x$ is occupied at time~$t$ and 0 elsewhen; the
sum over $v_{max}$ sites is necessary so that each bypassing vehicle is
really ``seen'' by the algorithm.  $n_T$ simply is the number of
vehicles which passed at~$x_0$ during the measurement time~$T$.

\changed
According to Fig.~2, our model reaches capacity ($=$ maximum
throughput) $q_{max} = 0.318 \pm 0.001$ at a density of $\rho^* :=
\rho(q_{max}) = 0.086 \pm 0.002$ for large systems ($L \ge 10^4$).
In addition, the figure shows that for a smaller system ($L = 10^2$)
the maximum throughput is much higher.  This means that short segments
behave differently from long ones!

\bfig
\epsfbox{../alife_archive/graph/fdiag.ps}
\fff .
Fundamental diagram of the model (throughput versus density).
Triangles: Averages over short times (200~iterations) in a
sufficiently large system ($L = 10,000$).  Solid line: Long time
averages ($10^6$~iterations) in a large system ($L = 10,000$).  Dashed
line: Long time averages ($10^6$~iterations) for a small
system~$L=100$.

\efig

A comparison with real traffic measurements~\cite{NaS92} indicates that
it is reasonable to assume that, at least to the order of magnitude, one
site occupies about $7.5~m$ (which is the space one car occupies in a
jam), one iteration is equivalent to about $1$~second, and maximum
velocity $5$ corresponds to about $120$ km/h.

It should be noted that the model so far can be treated
analytically~\cite{SchS93}.  The analytical results, however, are more
difficult to obtain; and the analytic methodology is not extendable to
more complicated situations like multi-lane traffic, ramps, or networks
(\cite{Rickert1,Rickert2}, and see below).

Other work using (mostly even simpler) cellular automata for traffic
flow on roads is, e.g.,~\cite{Nagatani_2_lane,Vilar_Souza}.

\subsec 2.2~Critical life-times of traffic jams

Looking closer at traffic pattern near capacity (as in Fig.~1), one
makes at least two observations:\bi

\item
Already at densities lower than $\rho^*$ (left column of Fig.~1), the
system displays spontaneous jams.  They are sometimes very rare: In
Fig.~1 their existence only shows up in the bottom picture of the left
column, near the right of the picture.

\item
Space-time plots of systems near $\rho^*$ have a remote resemblance to a
directed percolation transition~\cite{DP} and the emergence of the giant
component in random graphs~\cite{Bollobas} in the way that jams at
densities $\rho^*$ have a finite life-time, whereas there seem to be
jams of infinite life-times and spanning jam-clusters above~$\rho^*$
(right column in Fig.~1).

\ei
A jam-cluster is roughly defined in the following way: Spontaneous
formation of a jam is caused by one car accidently coming too close to
the one ahead of it, which leads to a lower speed than normal.  Other
cars which have to slow down because of this car are ``in the same
jam''.  The life-time~$T_{life}$ of this jam-cluster is the time until
this structure is dissolved (i.e.\ no more cars with speed lower than
normal).

\changed
In a more quantitative treatment, we measure the distribution of jam
life-times using closed systems with different densities.  We plot the
number $N(\ge t)$ of jams with a life-time longer than $t$ as a
function of~$t$ in a doublelogarithmic plot~\cite{Nag94}.  (Technically,
\[
N(\ge t) := \sum_{\tilde t = t}^\infty n(\tilde t) \ ,
\]
where $n(\tilde t)$ is the number of jams with a life-time exactly $=
\tilde t$ in a given simulation run or number of runs.)
For a true percolation-like transition~\cite{Stauffer_Aharony} one
would expect a behavior as depicted in Fig.~3a.  Roughly speaking, the
curves mean that, at low densities, long life-times are very
improbable.  However, at densities higher than critical, the system
should be dominated by a few ``very longlived'' jams, which only leave
room for shortlived jams between them.  This leads to the $N(\ge
t)$-curve becoming horizontal for large $t$.  And in between one would
expect a ``critical'' density, at which these curves converge towards
a straight line ($N(\ge t) \propto t^{-\alpha}$) for $\rho \to \rho_c$
and $t \to
\infty$.

\bfig
\epsfbox{../alife_archive/graph/ltime_theo.ps}

\epsfxsize\hsize
\epsfbox{../alife_archive/graph/ltime.ps}

\fff .
Theoretical ({\it top\/}) and simulated ({\it bottom\/}) distribution
functions of life-times of traffic jams for a model where the
spontaneous initiation of jams is impossible ($p_{spont} = 0$).  The
curves show, for different densities, the number of jams with a
life-time larger or equal than $t$ as a function of~$t$.  (y-axis
arbitrary units)

\efig

%
%
%
%

In practice, we find a more complicated behavior for our
system~\cite{Nag94,NPB94}. The following is a short interpretation:\bi

\item
The model has a certain probability $p_{spont}$ of the spontaneous
initiation of a new jam, which depends on the density of cars and on the
amount of fluctuations which happen when cars move at full speed.

\item
This probability provides an upper cut-off on the length-scale and on
the time-scale, up to which the model can display critical behavior.

\ei
\changed
``True'' critical behavior can be recovered, when the model is redefined
in a way that $p_{spont} = 0$.  In terms of the model, this means that
one has to reduce the fluctuations of free driving (i.e.\ undisturbed by
other cars) to zero.  Note that this leaves the fluctuations at
accelerating and at slowing down unchanged.  Once all cars have reached
maximum speed (if density allows that), no new jam may initiate itself
spontaneously.  One can then externally initiate one jam at a time
(e.g.\ by picking on car randomly and reducing its speed by 1) and
measure the properties of this jam.  Doing this with different
densities, we obtain the results of Fig.~3 bottom, which show that in
this particular limit, the model corresponds indeed closely to the
theoretical picture.

Please refer to the above-mentioned references for a more complete
description.

\subsec 2.3~Variability and predictability of travel times

Measuring the life-time distribution of traffic jams is convenient for a
theoretical understanding, but it is not very useful for everyday
traffic.  The probably most important reason for this is that life-times
of jam-clusters are practically not amenable to measurements.

A quantity which is much easier to measure and which is extremely
relevant in the context of transportation management is the individual
travel-time and its variation from vehicle to vehicle using the same
route.  For the following simulations, we still use a closed
loop of size~$L$.  We define a subsegment of length~$l<L$ and
measure, for each car, the time~$t_l$ between entry and exit of this
subsegment.

The relative variation of travel-times is defined as
\[
\sigma(t_l) := { \sqrt{ \mede{ ( t_l - \mede{t_l} )^2 } } \over  t_l } \ .
\]
$\mede{\ldots}$ denotes the average over all cars during the
simulation; $\mede{t_l}$ therefore is the average travel-time for all
cars during the simulation.

Results of these measurements as a function of density are shown in
Fig.~4.  We use a system of length $L = 10^3$ and measure trips along a
designated subsegment of $l=100$.  The simulation runs for $10^5$~time
steps, and every time a car finishes a complete travel along the
measurement subsegment, its travel time is taken into account for the
average.

\bfig
\epsfbox{../alife_archive/graph/ttime.ps}

\epsfxsize\hsize
\epsfbox{../alife_archive/graph/sigma.ps}

\fff .
Travel time and variations of travel time as a function of density.
System size~$L = 10^3$, length of traveled subsection~$l = 10^2$,
measured time $T = 10^5$ time-steps.

\efig

One clearly sees that both the travel time and the vehicle-to-vehicle
fluctuations are approximately constant up to a density around 0.09.
There, the travel time starts to rise as a function of density, and the
fluctuations go up very steeply and reach a maximum near $\rho = 0.11$.
In other words, one can not only show that the region of maximum
throughput shows near-critical behavior in a theoretical sense, but also
that this behavior has practical consequences: It implies that, passing
from slightly below to slightly above capacity, one comes from a regime
where the travel time is predictable with an accuracy of approx.~$\pm
3\%$ to a regime where the error climbs up to $\pm 65\%$ or more.

\subsec 2.4~Traffic at the edge of chaos

Summing up our results, we obtain the following picture: Near maximum
throughput, our model shows scaling of jam life-times and high
variability of travel time, features which indicate a critical phase
transition.  But the life-time scaling shows an upper cut-off; and the
density of maximum throughput does not exactly coincide with the
density of maximum fluctuations.  Thus, the transition is not truly
critical (although we use the word criticality throughout the text).
However, it becomes exactly critical in the limit of zero fluctuations
for free driving (i.e.\ not influenced by other cars).

A helpful concept for understanding critical phase transitions in
discrete systems is the notion of ``damage
spreading''~\cite{damage_spreading}: One simulates two identical copies
of the system.  At a certain point, a minimal change in one of the
copies is made and then the time evolution of the {\em differences\/}
between the systems is observed.

In our traffic system, ``damaging'' means to change the velocity of one
randomly picked car by $-1$.  This car then causes a jam of a certain
life-time; and downstream of this jam, the traffic pattern will be
different from the undisturbed model.  After this jam has dissolved, the
{\em spatial\/} amount of damage extends from the disturbed car to the
last car involved in the jam, and this length is proportional to the
life-time of the jam.  For the limit $p_{spont}
\to 0$ (but $p_{spont} \ne 0$),
i.e.\ where spontaneous initiation of a jam becomes rare, one obtains
the following picture:\bi

\item
For low densities $\rho << \rho_c$, jams are usually short-lived
(i.e.\ with an exponential cut-off in the life-time distribution).  As
a result, the average amount of spatial damage is small.

\item
When approaching the critical density $\rho_c$, jams become
increasingly long-lived, with the result that the amount of spatial
damage becomes larger and larger.  Ultimately, exactly at the critical
point, a damage of size infinity (in the thermodynamic limit) is
possible.

\item
Above the critical point ($\rho > \rho_c$), the jam caused by the
disturbance will (in the average) survive forever, thus (in the
average) causing infinite damage.

However, traffic for $\rho >> \rho_c$ is characterized by the
existence of many jams quasi-randomly distributed over the system.  So
the additional jam caused by the disturbance will not change the {\em
statistical\/} properties of the system.

\ei
All these observations are si\-mi\-lar to con\-ven\-tio\-nal damage
spreading observations in cellular automata~(CA)~\cite{Langton92}: The
damage is limited for class~I and class~II CA, and it {\em can\/} be
infinite for class~VI CA.  For class~III CA, the damage is practically
always infinite, but does not change the statistical properties of the
system.

In summary, for the limit $p_{spont} \to 0$, $p_{spont} \ne 0$ we have
in our probabilistic CA a phase transition of the traffic patterns
si\-mi\-lar to the one found in more conventional and determinstic CA.
The control parameter in our case is the density, whereas in CA rule
space it is still an open question how to derive an order parameter
from the rules~\cite{Langton92,Italiener}, or if this is at all
possible~\cite{Mitchell}.

But, as stated initially in this section, a more realistic version of
the model with $p_{spont}$ significantly different from zero moves the
point of maximum throughput away from the critical point.  Therefore,
we have, similar to other
systems~\cite{earthquakes%
}, the existence of traffic in the {\em vicinity\/} of ``the edge of
chaos''.

The rest of this paper will be devoted to arguing why this regime is
of special importance for transportation.

\sec 3.~A simple transport network

We now move away from the single lane closed loop system to a single
lane highway network with ramps connecting the different segments.
The travelers on this network have route plans so that they know which
ramps they need to exit to reach their individual destinations.  We
assume that each traveler always has the same origin/destination
pair.  Each traveler remembers the last travel time for each
alternative route between his or her origin and destination.  The
network may have traffic density sensors at specified locations which
can be used to identify congested areas and perhaps introduce toll for
the use of such links.  The travelers are able to re-plan depending on
their aggregated transportation costs which is their remembered travel
time plus eventual toll.  Such a sensor setup is an example of a
(Advanced) Transportation Informational System (ATIS), and the
introduction of toll for the use of highly congested links is a simple
example of an (Advanced) Transportation {\em Management\/} System
(ATMS)~\cite{ATIS,IVHS}.  The rationale behind such a toll policy is
to make the highway traffic more efficient by pushing a larger part of
the system towards the density corresponding to maximum flow.
Interestingly, this implies that more traffic intentionally will be
moved into the critical regime as defined above which in turn will
increase the fluctuations of the travel times as well as the
non-predictability of transportation system dynamics.  This effect is
the topic of this section.

Some attempts have been made earlier to use CA techniques to simulate
simple representations of network traffic.  Most of the models map
network traffic on particles hopping on a 2-D square
grid~\cite{Biham,Nagatani_2_level,Cuesta}.  These models are very
useful to understand the transition from a free flowing to a
jamming phase in urban traffic, but in these models the maximum flow
of traffic is given entirely by the capacities of the
intersections~\cite{Nagatani_2_level}, which is not always realistic,
e.g.\ for arterials.  Closest to our approach is~\cite{Schuett}, which
however was never used in order to do systematic studies like the one
presented here.

\subsec 3.1~Ramps

In order to simulate this simple network, we first need reasonable
algorithms for transferring vehicles from one road to another at
junctions.  This involves two parts: Including the vehicle into the
traffic stream on the target road; and then deleting it from the
source road.

Unfortunately, introducing an additional car into a given traffic
stream can cause some problems.  Just adding the ramp-inflow to the
traffic on the main road easily leads to disturbances which (i) block
the traffic on the main road, and (ii) lead to an outflow, downstream
from the ramp, which is {\em below\/} capacity.  For this reason, we
chose an algorithm where access to the main road is only possibly when
there is sufficient space between vehicles.  We believe that this is
realistic enough to represent metered ramps (i.e.\ ramps with
regulated access), and since we are often concerned with the analysis
of future traffic systems, it seems appropriate to model a technically
advanced traffic control system here.

The algorithm works as follows.  Imagine a ramp, as in common
experience, as two parallel stretches of road; these parallel
stretches have a length of 5~sites in the model.  The target stretch
is part of a longer road and therefore is connected at both ends,
whereas the source stretch is only backwards connected.  If there is a
vehicle (velocity~$v$) on the source stretch, then\bi

\item
it looks, on the target stretch, for the next car ahead (which may be
its neighbor; $\leadsto gap_{forward}$); and

\item
it looks for the next car behind on the target stretch ($\leadsto
gap_{backwards}$).

\item
Then the following rules are applied:
\bpgm
\qquad IF~($gap_{forward} > v$ .AND. $gap_{backwards} > v_{max}$ )\cr
\qquad\qquad change\_lane \cr
\qquad\qquad $v = \max( v_{max}, gap_{forward} )$ on new lane \cr
\qquad ELSE \cr
\qquad\qquad IF ( $v \ge 1$ ) go\_one\_site\_backwards~~~~~~$(*)$\cr
\qquad\qquad $v = 0$ \cr
\qquad ENDIF \cr
\epgm

\ei
One may imagine that this is emulating a ramp metering system, where a
technical device upstream of the ramp determines where to fit in a
car.  The car then gets a green light and arrives at maximum speed, in
between two other cars on the target road.

Line $(*)$ is a technicality.  It is necessary because the global
velocity update may reaccelerate the speed of the vehicle to one and
then move it one site ahead.  If the car repeatedly fails to change to
the target lane, then the car would slowly advance on the changing
area and ultimately leave it.

The details of this algorithm will probably not matter for our results,
as long as it allows maximum flow downstream from the ramp.  That this
indeed is the case is shown in Fig.~5, which may be compared to Fig.~1.
It gives the fundamental diagram for a system with two road segments
where one is a closed loop and the other one provides an alternative
route for a certain length, connected to the main road by one exit and
one entry ramp.  Half of the vehicles use this alternate lane.  Density
and throughput are measured on the undivided part.

\bfig

\epsfbox{../alife_archive/graph/ramp/ramp.ps}

\fff .
Fundamental diagram for ramp.  A circular segment of length~$L=10^3$
may be partially bypassed by a second segment.  50\% of the traffic
uses the bypass; at the end of the bypass, it again merges with the
main stream of the traffic.  The measurements were taken at the part
of the main segment where no bypass exists.

\efig

\subsec 3.2~Nash Equilibrium versus System Optimum

An important issue in the context of a transportation network is the
difference between Nash Equilibrium (NE; $=$ User Equilibrium, UE) and
System Optimum (SO); the optimum dynamics of an individual traveler
versus the situation where the capacity of the transportation system
is used in the most optimal manner.  These two systems states are
often in conflict.

This conflict can perhaps best be illustrated in terms of a simple
transport network example (a variation of~\cite{Catoni}), especially
as we will use the same example for simulation experiments later on.

Imagine a road from A to B with capacity~$q_{max}$, with a bottleneck
with capacity~$q_{bn}$ shortly before B.  (Fig.~6 shows the same
network, albeit different traffic patterns.)  Further imagine that
there exists an alternative, but longer route between A and B.  On the
direct route from A to B additional travelers from C have to go to
destination D.  First assume that there are no travelers with origin
in C.

\bfigtwo
$$
\hbox{
\vbox{{
\baselineskip5pt{
\vipt
\begin{verbatim}
10:
====================================================================================================================================
                   ....................................... //
.................................................................
                   .                            (alternative route)
                                              .
                   .
                                              .
                   .
                                              .
                   .
                                              .
                   .
                                              .
                   .
                                              .
 (A)          #.....                              (direct route)
                         V1V1V1V1V1            (B)
#####04.....5.....3....3.....4.....1.4......5......5..........................................................................#####
                                  .....#
      #.....
                                  .
           .
                                  0
           .
                                  0
           .
                                  .
           .
       (C)                        .
           .                             (D)
      #####0.3....4.....4.....2...1
           .............................#####
30:
====================================================================================================================================
                   ................4...................... //
.................................................................
                   .
                                              .
                   .
                                              .
                   .
                                              .
                   .
                                              .
                   .
                                              .
                   .
                                              .
 (A)          #.....
                         V1V1V1V1V1            (B)
#####1.2...1.01..2..1...2..3.....5.......4......5......5...........5.......4.......4.....4......4..........0..1...1...........#####
                                  .....#
      #.....
                                  .
           .
                                  .
           .
                                  0
           .
                                  0
           .
       (C)                        0
           .                             (D)
      #####0.01.1..0.01.01.00.00000
           ...................5.........#####
210:
====================================================================================================================================
                   .......5............................... //
............4.........5..........................................
                   .
                                              .
                   .
                                              .
                   .
                                              .
                   .
                                              .
                   .
                                              .
                   4
                                              .
 (A)          #4....                                             (*)
                         V1V1V1V1V1            (B)
#####0.4....4.........4................4......2...2...1.2....000.000000.1...3.......4.........4........3.....1..1..2..........#####
                                  .....#
      #.....
                                  .
           .
                                  .
           .
                                  2
           .
                                  .
           .
       (C)                        1
           .                             (D)
      #####03....01.1.1.01.0000001.
           ........................4....#####
\end{verbatim}
}
}
}
}
$$

\fff .
Schematic network representation with traffic, showing time-steps~10,
20, and 210 of a particular simulation.  Traffic entering at (A) is
bound for (B) and may use the ``direct route'' or the ``alternative
route''.  Traffic entering at (C) is bound for (D).  The bottleneck is
denoted by {\tt V1V1V1V1V1} (maximum speed 1).  One observes that the
traffic coming from (C) has difficulties entering into the main
stream; and---in time-step~210---a disturbance denoted by {\tt (*)}
has traveled backwards from the bottleneck.

\efigtwo

If many drivers are heading from A to B, they will, without knowing
anything about the overall traffic situation, all enter the direct road.
In consequence, a queue builds up from the bottleneck.

A Nash-Equilibrium is defined as a situation where no agent ($=$
driver) can lower his or her cost ($=$ decrease travel time) by
unilaterally changing behavior.  Assuming that the drivers have
complete information, this implies that the waiting time in the queue
exactly compensates for the additional driving time on the alternative
route.

Now assume that there are additional travel demands from C to D (see
Fig.~6), the exit for the latter lying shortly before the bottleneck.
Obviously, this traffic is suffering from the bottleneck queue upstream
($=$ left) of the bottleneck, and from these travelers point of view it
would be much better if the queue were located to the left of the ramp
that the travelers from C use to enter the link.  Note that moving the
queue further upstream does not make any difference for the drivers
originating in A.

This example illustrates that one easily finds situations where there
are better overall solutions than the NE.  (Technically, a SO is reached
when the sum of all individual costs ($=$ travel times) is minimal.)
Recent simulation results~\cite{Mahm_SO_NE} indicate that the SO could
give performance advantages of about $15\%$ for realistic situations.

A way to push a traffic system from a NE towards a SO is to keep the
density on each road at or below $\rho^*$, the density of maximum
throughput.  Then there would not exist queues anywhere in the system,
thus ensuring that additional traffic could proceed undisturbed.  Note
that this could for instance imply (in the limit of a perfect
implementation) that drivers have to wait to enter the road network
until sufficient capacity is available for them.

\subsec 3.3~Travel plans and individual decision logic

In our simple network, there are only two different types of travelers:
Travelers from A to B, and travelers from C to D.  Travelers from A to B
can choose between the direct and the longer alternate route.  In order
to make decisions, each AB-driver remembers his or her last travel-time
on each of the two routes.

A traveler calculates expected costs~\cite{Arnott} according to
\[
cost_{direct} = toll + \alpha \cdot t_{direct}
\]
and
\[
cost_{alt} = \alpha \cdot t_{alt}
\]
where $cost_{direct}$ and $cost_{alt}$ are the expected costs for the
two route choices, $toll$ is the toll for the current day (see below),
$t_{direct}$ and $t_{alt}$ are the remembered travel times for each
route, and $\alpha$ is a conversion factor which reflects trade-off
between time and money.  $\alpha$ could be different for each driver,
but is uniformly equal to one in this work.  ($\alpha$ reflects
``standard values of time'', VOT, which can be looked up for traffic
systems.)

Then, each driver chooses the cheaper route, except that there is a
$5\%$ probability of error (which gives each driver a chance from time
to time to update her information about the other possibility).

As long as the traffic dynamics is deterministic and completely
uniform, this scheme leads to a Nash equilibrium~\cite{Arnott}.
However, in our case of stochastic traffic dynamics, this is no longer
true: There might well be a decision rule different from the one above
where at least one traveler is better of, for example by triggering
from some kind of day-to-day oscillation between the two routes and
taking advantage of it.  In other words, by dealing with stochastic
traffic dynamics, the notions of economic equilibrium theory have to
be used with care.

\subsec 3.4~Space-time dynamics

Before we discuss how to determine the toll, we shortly turn to a
space-time plot of the direct route from A to B (Fig.~7).  As in each
part of Fig.~2, vehicle movement is to the right and time is
downward.  The figure contains the first 300~time-steps, and then
time-steps~2000 to 2950.

\bfigtwo

\advance\epsfxsize by-1in
\centerline{\epsfbox{../alife_archive/graph/seg1.ps}}

\fff .
Space-time plot of the main segment (A-B) of the network. The cars are
injected at the left.  About one half inch to the right, a change in
gray indicates the junction where vehicles to the alternative route
leave.  Another one and a half inches to the right, the jam structures
indicate the on-ramp for travel from C.  About one inch from the
right, another change in gray indicates the off-ramp to D (see arrow
on top).  Very close to the right is the bottleneck, together with
jams emerging from this region and traveling backwards into the
system.

\efigtwo

The major dimensions of the system are:\bi

\item direct route from A to B: 1021 sites (full size of plot)

\item cars leave for the alternative route at position~111

\item cars coming from C enter at position~322 and leave again at
position~881

\item the bottleneck ($v_{max}=1$) extends from position 1001 to 1011.

\ei
$20\%$ of the A-B vehicles are preselected to leave at the junction
for the longer route, as can be seen in the picture by a change of the
gray shading.  The entry-point of the C-D vehicles is marked by the
permanent existence of a disturbance, which is very often connected to
other disturbances which travel ``backwards'' through the system.

The point of exit for the C-D vehicles is covered by dense traffic
most of the time, but it may be seen near the top right of the figure
as a change in gray shading and as a sudden stop of some
trajectories.

The bottleneck is visible at the very right edge of the figure, where
the trajectories of the vehicles are diagonally pointing downwards to
the right.

The striking feature of this picture is the graphic illustration of
the highly dynamic and (seemingly) nonlinear structure of traffic
patterns.  Vehicles do {\em not\/} wait orderly in front of the
bottleneck, but instead self-organize into backwards moving jam waves.
If one of these waves reaches back into an area with higher density
(in our case the junction where the C-D cars leave), then the survival
probability of this jam wave suddenly becomes much larger, and it may
move deeply back into the system.  A single snap-shot of such a
traffic situation could not uncover the origin of such a wave.  The
implications to traffic measurement and modeling are important.

Something similar is true for the region where the C-D traffic stream
enters the main road.  It is not a process where both traffic streams
line up to wait until they can jointly proceed.  Instead, it is often
even possible that the additional traffic enters into the main stream
without causing a major disturbance right away.  But the locally
enhanced densities is unstable and leads sooner or later to the
initiation of a disturbance, which then travels backwards to the
junction (and often beyond).

These results indicate that the methodology of queueing
networks~\cite{Powell} has to be handled with care when applied to
vehicular traffic.

\subsec 3.5~Congestion detection, toll and travel pricing

\changed
For simplicity, assume that the current toll is based on some traffic
observation on the last period (day).  Let us further assume that each
driver only drives this trip once in each period (day).  (Note that
this is oversimplistic, and further investigations are needed to make
it work for, e.g., workdays versus weekend-days.)

Algorithmically, we can proceed as follows: (i)~The traffic
microsimulation is executed for one period.  Each driver updates her
travel time information just after arrival at the destination.
(ii)~After all cars have reached their destinations, the toll is
adapted according to the average value provided by the sensor.
(iii)~Each A-B driver makes her route choice.  (iv)~The next
microsimulation period starts.  --- This results in a day-to-day
evolution of the decision pattern~\cite{Mahm_C_Herman}.  The procedure
is actually very similar to standard game-theory~\cite{Axelrod},
except that we obtain the pay-off from the microsimulation and not
from a predefined matrix.

\changed
A critical question remains: Where should one place the traffic
sensor(s) for the determination of the toll?  Placing it {\em
inside\/} the bottleneck is not very useful, because traffic there is
always at or below the ``efficient'' density (i.e.\ at or below the
density corresponding to maximum throughput).

\changed
Intuitively, it would make sense to measure the length of the queue in
front of the bottleneck.  However, as we showed in the last section,
the dynamics of the traffic does not lead to the built-up of a regular
queue but to a system of backtraveling jams instead, which makes this
approach infeasible.

\changed
Therefore, we chose to measure the average density on the segment
upstream of the bottleneck, i.e.\ between the exit to D and the
bottleneck.  Then, the next question is, which should be the target
density for the control algorithm?  When one is measuring traffic
upstream of a bottleneck~\cite{Hall}, then stationary traffic can
never reach maximum throughput: Either traffic operates at densities
corresponding to flow rates lower than the bottleneck capacity, or
dense traffic builds up.  Traffic can only ``use'' the part of the
fundamental diagram which is below the capacity of the bottleneck; in
consequence, densities are either far below or far above the ones
corresponding to maximum throughput.

\changed
However, having some knowledge about the bottleneck is not really
helpful: In a more complicated traffic network, it may be the case
that further downstream from one bottleneck there is another one,
which has even lower capacity.  Or the bottleneck may be the on-ramp
to a crowded major road: Here the performance of the bottleneck
depends on the time-dependent and fluctuating load on the main road.

\changed
We therefore follow a simplistic and completely local approach here,
which will nevertheless prove to be quite effective.  Assume that the
toll is operated by a local ``toll agent'', who does not have any
global knowledge.  However, she knows the fundamental diagram (flow as
function of density) of {\em her\/} sensor area.  If she wants to keep
the traffic at maximum flow, she has to keep the density in the
correct range, i.e.\ near~$\rho = 0.08$.  We implement this by the
following rules:
\bpgm
\qquad IF ( $\rho < 0.06$ ) THEN \cr
\qquad\qquad toll = toll - delta \cr
\qquad ELSE IF ( $\rho > 0.10$ ) THEN \cr
\qquad\qquad toll = toll + delta \cr
\qquad ENDIF \ , \cr
\epgm
where {\it delta\/} is an external parameter.

\changed
According to our arguments above, is not obvious that this approach
will produce meaningful decision behavior: The toll agent tries to
keep the traffic at a density regime which is dynamically impossible
because of the bottleneck downstream.  It is not clear, a-priori, what
effect this will have, and it was one of our main points of interest
in how this control mechanism would work.

\sec 4.~More simulation results: How to play traffic games

\subsec 4.1~Technical Set-up

In the following, we describe one particular simulation run in more
detail.  We used a network of overall size 1962~sites, composed of the
following parts:\bi

\item direct route from A to B: 241 sites (smaller than for Fig.~7 to
reduce computational demand)

\item
alternate route from A to B: 1570 sites ({\em much\/} longer than
direct)

\item connection from C to main route: 103 sites

\item connection from main route to D: 48 sites

\item length of the section shared by A-B and C-D-travelers: 101
sites

\item thus, overall length from C to D: 252 sites

\item length of bottleneck (with maximum velocity reduced to one): 10
sites

\item position of the bottleneck: starts 20 sites before reaching B

\ei
The density $\rho_{toll}$ for the update of the toll is measured
between the junction where the vehicles heading for D leave the main
route, and the start of the bottleneck.

We have $N_{AB} = 16000$ vehicles which want to travel from A to B, and
the same amount $N_{CD} = 16000$ which want to travel from C to D.  At
each ``day'' of the simulation, they are lined up outside the
simulated system in the same sequence; and they enter the system at
their respective entry points as soon as the simulated traffic allows
it (cf.~Fig.~6).

When the vehicles enter the system, they already have decided on their
travel plans, so they just execute these plans.  The simulation runs
until all vehicles have reached their respective destinations.  Then
the toll is updated and drivers decide their route for the next day,
as described above.

\subsec 4.2~A simulation of 200 periods (days)

We describe 200 days of a simulation where the toll was kept at zero
during the first 100~days, and in addition all A-B-travelers were
forced to use the direct route during the first 50 days.

Fig.~8a shows results for the trip times and the adaptive toll, Fig.~8b
the vehicle-to-vehicle variations of the trip time (as defined earlier),
and Fig.~8c the day-averaged density, on selected road sections.  These
sections are: (i)~the section where the density for the toll adaption is
measured, (ii)~the section of the main road between the on-ramp from C
and the off-ramp from D, and (iii)~the alternative route.

\bfig

\epsfbox{../alife_archive/graph/50_50/net_ttimes.ps}

\vskip3mm

\epsfxsize\hsize
\epsfbox{../alife_archive/graph/50_50/net_sigma.ps}

\vskip3mm

\epsfxsize\hsize
\epsfbox{../alife_archive/graph/50_50/net_dens.ps}

\fff .
Simulation output for 200~iteration of the simple corridor network
model.  Time-steps 1-50: No adaption; 51-100: drivers can choose
alternative route; 101-200: drivers can choose alternative route, and
the toll adapts in order to keep the density at the specified level.
{\it Top:\/} Average trip times for the direct and for the alternative
route from A to B as well as for the route from C to D, and toll for
the direct route from A to B.  {\it Middle:\/} Vehicle-to-vehicle
fluctuations of trip time for the direct and for the alternative route
from A to B.  {\it Bottom:\/} Densities on the segment shared by
A-B-direct travelers and C-D-travelers, on the segment shortly before
the bottleneck used for determination of the toll, and on the
alternative route from A to B.

\efig

Even when allowed (i.e.\ after day~50), not many of the A-B drivers use
the new option of the alternative route.  This is to be expected, since
it is more than six times longer than the direct one.  In consequence,
travel times and fluctuations do not change much.

After day 100, the adaptive tolling starts and fairly quickly reaches a
stationary value around 260.  As the ``toll'' line in Fig.~7c indicates,
this keeps indeed $\rho_{toll}$ near the specified range between
$\rho=0.06$ and $0.10$.  In addition, the density on the main segment
(used by both A-B and C-D travelers) drops to around $0.11$, above, but
close to the density of maximum throughput.

Travel times for C-D and for A-B-direct travelers go down (Fig.~8a);
and the toll just offsets the time gain for use of the direct route:
$time_{direct} + \alpha \cdot toll \approx time_{alternat.}$; remember
that $\alpha = 1$.

Vehicle-to-vehicle fluctuations (Fig.~8b) for the use of the alternative
road go up from ca.~2\% to around~12\%, and for the use of the direct
road from ca.~11\% to around~42\%.  Moreover, the {\em day-to-day\/}
fluctuations also seem to go up in all measurements.

All this is in agreement with our intuition that traffic management
can indeed make traffic more efficient, but may in addition lead to
higher fluctuations and, as a consequence, lower predictability, since
the system is driven closer to capacity and thus to the edge of chaos.

\changed
One should distinguish between two different kinds of fluctuations:
Fluctuations due to the dynamics, and fluctuations due to the
learning. The fluctuations in the latter might be due to the specifics
of the chosen learning scheme, especially the lack of historic
information beyond the last day.  More realistic assumptions about the
learning and en-route information are claimed to avoid
that~\cite{k.axhausen}.  However, the results for the
vehicle-to-vehicle fluctuations (i.e.\ the $\sigma$ as defined in the
text) only depend on the fact that the traffic density is driven
towards the critical value.  A less fluctuating learning scheme should
therefore even {\em increase\/} our values for $\sigma$.


\sec 5.~Conclusion

We started out establishing/reviewing some facts for a simple
closed-loop single-lane system:\bi

\item
Traffic at maximum capacity is in a regime which is critical up to an
upper cut-off.

\item
This upper cut-off depends on the probability~$p_{spont}$ for the
spontaneous initiation of a jam.

\item
The predictability of travel times sharply decreases when the density
goes above this point.

\ei
This leads to the observation that advanced flow control will not only
affect traffic flow, but will moreover drive large portions of the
system towards the critical regime.  The main reason for this is that
the most efficient use of a traffic system takes place when all parts
operate at densities at or below capacity.  Systems designed for the
management of traffic flows will reroute traffic from overcrowded
roads to undercrowded ones, thus driving both closer to criticality.
(We use criticality in this text even for the ``not truly'' critical
situations, as discussed in the text.)  Once traffic is near the
critical region, further control inputs will have very unpredictable
consequences.

More precisely, the following occurs.  If one assumes complete
information and rational decisions by everybody, the traffic will aim
towards a Nash Equilibrium (NE).  As nowadays drivers do not have
complete information, we assume that they do something like bias their
decisions towards ``safe'' routes, preferring e.g.\ shorter routes
over longer ones even if both yield the same travel time.

Advanced Traveler Information Systems (ATIS)~\cite{ATIS,IVHS} are
developed to enhance the amount of information available.  As
explained above, this will push the traffic system closer to the NE
and therefore---because it spreads traffic out over the
network---closer to criticality.

Moreover, traffic management will aim beyond the NE towards a System
Optimum (SO).  A necessary condition for a SO is that no part of it is
operating above the density $\rho^*$ of maximum throughput, which will
drive the system again closer to criticality.

\changed
This implies that the approximation of deterministic, predictable
traffic patterns would be less and less correct the more one
approaches high performance of the traffic system.  In consequence,
traffic assignment methods based on relaxation to equilibrium would no
longer be meaningful: The changes in the traffic patterns due to one
relaxation step would get lost in the changes due to the inherently
fluctuating dynamics, and the algorithm would never converge.  An open
question is inhowfar one can replace the equilibrium quantities by
statistical averages (e.g.\ many Monte-Carlo runs); this is a topic of
future research.

One envisaged way~\cite{Mahm_SO_NE} of reaching the SO is to give each
driver individual route guidance instead of complete traffic
information.  If one doubts that this will lead to high user
acceptance, then congestion pricing seems to be the only alternative.
Our simulation results support the idea that already locally operating
agents can achieve this in an efficient manner.

\changed
In the text, we discuss the case of tolling on a specific road
segment upstream of ``the'' bottleneck. This demands prior knowledge
about the system.

However, one can imagine a completely local algorithm in the following
way (see also~\cite{varian}): Assume that {\em every\/} road segment
in the system is operated by a simple economic agent.  This agent
wants to keep the operation of the segment as efficient as possible,
and the only measure she has is to go up or down with the toll.  The
agent knows the performance characteristics (i.e.\ throughput $q$ as a
function of density~$\rho$) of her segment, and from this she obtains
the density which corresponds to maximum flow and therefore to maximum
road performance.  The agent then tries to keep the density on her
segment at this particular density, increasing the toll when the
density becomes too high, and else decreasing it.  In a real network,
we would expect that the toll for most segments turns out to be zero.

This tolling scheme gives the impression that every agent locally
drives her segment towards criticality ($=$ maximum flow), but the
situation is more complicated.  In most cases, it is not the traffic
{\em inside\/} bottlenecks which is tolled, but the overcrowded
segments {\em upstream of\/} the bottlenecks.  But because of the
bottleneck, these upstream segments usually cannot operate at maximum
throughput: As soon as the incoming flow is more than the bottleneck
capacity, dense traffic builds up, and the segment switches from
operation far below to far above the critical point (see text).
Nevertheless, our results show that this still leads to having more
parts of the network near criticality, as a result of collective
effects.

In an economic context, we therefore have a local aiming for high
performance, which happens to coincide with criticality.  But even
though the criticality very often cannot be reached locally by this
mechanism, it drives the {\em global\/} system closer towards
criticality: Local maximization of efficiency leads to global
criticality~\cite{SOC}.

Or in short: The fact that, in a complex system, high performance
often has the downside of high variability
seems also to be true in transportation systems.

\section*{Acknowledgments}

KN is also a member of the ``Graduiertenkolleg Scientific Computing
K\"oln--St.~Augustin''.  We thank S.R.~Nagel, D.~Stauffer, and others
for making us aware that phase transitions which are not ``truly''
critical are still interesting and maybe even more relevant for
reality.  Nick Vriend, Kay Axhausen, and the anonymous referees
provided helpful comments.  We are especially grateful to Chris
Barrett for carefully reading and commenting on the first draft.  ZPR
K{\"o}ln and TSA-DO/SA (LANL) provided most of the computer time.


\begin{thebibliography}{99}

\def\jrn#1{{\em #1}}
\def\rf#1 #2 #3.{#1:#3.}

\bibitem{Meyer_Miller84}
Meyer, M.D., and E.J.~Miller. 1984. \jrn{Urban Transportation
Planning}.  McGraw-Hill Series in Transportation.

\bibitem{Cohen}
Cohen, J., and F.~Kelly. 1990. A Paradox of Congestion in a Queuing
Network. \jrn{J.\ Appl.\ Probablilty\/} \rf 27 90 730-734.

\bibitem{Arnott}
Arnott, R., A.~De Palma, and R.~Lindsay. 1993. A structural model of
peak-period congestion: A traffic bottleneck with elastic demand.
\jrn{The American Economic Review\/} \rf 83(1) 93 161-179.

\bibitem{Alife4}
TRANSIMS, The TRansportation ANalysis and SIMulation System project at
the Los Alamos National Laboratory.

\bibitem{Rickert1}
Rickert, M. \jrn{Outflow from 2-lane traffic simulations.} In
preparation.

\bibitem{Rickert2}
Rickert, M. 1994.  \jrn{Simulationen zweispurigen Autobahnverkehrs auf
der Basis von Zellularautomaten}.  Master Thesis, Univ.\ of Cologne.

\bibitem{Wolfram}
Wolfram, S. 1986. \jrn{Theory and Applications of Cellular Automata}.
Singapore: World Scientific.

\bibitem{Stauffer}
Stauffer, D. 1991. Computer simulations of cellular automata. \jrn{J.\
Phys.\ A\/} \rf 24 91 909-927.

\bibitem{NaSchleich}
Nagel, K., and A.~Schleicher. 1994. Microscopic traffic modeling on
parallel high performance computers. \jrn{Parallel Comput.} \rf 20 94
125-146.

\bibitem{Landau_Lifshiz}
Landau, L.D., and E.M.\ Lifshitz. 1986.  Statistical physics, Course in
Theoretical Physics, Vol 5.  Oxford: Pergamon Press.

\bibitem{paramics}
Wylie, B.J.N, D.~McArthur, and M.D.~Brown. 1992. PARAMICS
parallelisation schemes. EPPC-PARAMICS-CT.10, Edinburgh.

\bibitem{netsim}
Rathi, A.K., and A.J.~Santiago. 1990. The new NETSIM simulation
program.  \jrn{Traffic Engineering + Control} May 1990, 317-320.

\bibitem{NaS92}
Nagel, K., and M.~Schreckenberg. 1992. A cellular automaton model for
freeway traffic. \jrn{J.\ Phys.~I France\/} \rf 2 92 2221.

\bibitem{SchS93}
Schadschneider, A., and M.~Schreckenberg. 1993. Cellular automaton
models and traffic flow, \jrn{J.\ Phys.\ A} 1993, \rf 26 93 L679.

\bibitem{Nagatani_2_lane}
Nagatani, T.  1993.  Self-organization and phase transition in
traffic-flow model of a two-lane roadway. \jrn{J.~Phys.~A} \rf 26 93
L781-L787.

\bibitem{Vilar_Souza}
Vilar, L.C.Q., and A.M.C.~de Souza. 1994.  Cellular Automata Models for
General Traffic Conditions on a Line. Preprint.

\bibitem{DP}
Kinzel, W.  1983.  Directed percolation. In \jrn{Percolation structures
and processes}, edited by G.~Deutscher, R.~Zallen, and
J.~Adler. A.~Hilger.

\bibitem{Bollobas}
Bollobas, B. 1985. \jrn{Random graphs.}  London: Academic Press.

\bibitem{Nag94}
Nagel, K.  1994.  Life-times of simulated traffic jams, \jrn{Int.\ J.\
Mod.\ Physics C}. In press.

\bibitem{Stauffer_Aharony}
Stauffer, D., and A.~Aharony.  1992.  \jrn{Introduction to percolation
theory}.  London: Taylor \& Francis.

\bibitem{NPB94}
Nagel, K., M.~Paczuski, P.~Bak.  1994.  In preparation.

\bibitem{damage_spreading}
Stanley, E.H., D.~Stauffer, J.~Kert\'esz, and H.J.~Herrmann.  1987.
Dynamics of spreading phenomena in two-dimensional Ising models.
\jrn{Phys.\ Rev.\ Lett.} \rf 95(20) 87 2326-2328.

\bibitem{Langton92}
Langton, C. G.  1992.  Life at the edge of chaos.  In \jrn{Artificial
Life II}, edited by C.~Langton et al.  Santa Fe Institute Studies in the
Science of Complexity, Vol.~10.  Redwood City, CA: Addison-Wesley.

\bibitem{Italiener}
Bagnoli, F., R.~Rechtman, and S.~Ruffo. 1992.  Damage spreading and
Lyapunov exponents in cellular automata. \jrn{Phys.\ Lett.\ A} \rf 172
92 34-38.

\bibitem{Mitchell}
Mitchell, M., J.P.~Crutchfield, P.T.~Hraber.  1994.  Dynamics,
computations, and the ``Edge of Chaos'': A re-examination.
In~\jrn{Integrative Themes}, edited by G.~Cowan, D.~Pines, and
D.~Melzner.  Santa Fe Institute Studies in the Sciences of Complexity,
Vol.~19, Reading, MA: Addison-Wesley.

\bibitem{earthquakes}
de Sousa Vieira, M., G.L.~Vasconcelos, and S.R.~Nagel. 1993. Dynamics of
spring-block models: Tuning to criticality. \jrn{Phys.\ Rev.\ E} \rf 47
93 R2221.

\bibitem{ATIS}
Hall, R.W.  1993.  Non-recurrent congestion: How big is the problem?
Are traveler information systems the solution?. \jrn{Transpn.\ Res.\
C}~\rf 1(1) 93 89.

\bibitem{IVHS}
IVHS AMERICA.  1993.  \jrn{Surface transportation: Mobility, technology,
and society.  Proceedings of the IVHS AMERICA 1993 annual meeting}.
Washington, D.C.: IVHS AMERICA.

\bibitem{Biham}
Biham, O., A.~Middleton, and D.~Levine.  1992.  Self-organization and a
dynamical transition in traffic-flow models. \jrn{Phys.\ Rev.\ A\/} \rf
46 92 R6124.

\bibitem{Nagatani_2_level}
Nagatani, T.  1993.  Jamming transition in the traffic-flow model with
two-level crossings.  \jrn{Phys.\ Rev.\ E} \rf 48(5) 93 3290-3294.

\bibitem{Cuesta}
Cuesta, J.A., F.C.~Mart\'{\i}nez, J.M.~Molera, and A.~S\'anchez.  1993.
Phase transitions in two-dimensional traffic flow models.  \jrn{Phys.\
Rev.\ E} \rf 48(N6) 93 R4175-R4178.

\bibitem{Schuett}
Sch\"utt, H.  1991.  \jrn{Entwicklung und Erprobung eines sehr
schnellen, bitorientierten Verkehrs\-si\-mu\-la\-tions\-sy\-stems f\"ur
Stra{\ss}ennetze}.  Schriftenreihe der AG Automatisierungstechnik TU
Hamburg-Harburg No.~6.

\bibitem{Catoni}
Catoni, S., and S.~Pallottino.  1991.  Traffic Equilibrium Paradoxes.
\jrn{Transp.\ Sc.} \rf 25 91 240-244.

\bibitem{Mahm_SO_NE}
Mahmassani, H.S., and S.~Peeta.  1993.  Network performance under system
optimal and user equilibrium assignments: Implications for Advanced
Traveler Information Systems. \jrn{Transportation Research Record} \rf
1408 93 83-93.

\bibitem{Powell}
Sim\~ao, H.P., and W.B.~Powell.  1992.  Numerical methods for simulating
transient, stochastic queueing networks. \jrn{Transpn.\ Sci.} \rf 26 92
296.

\bibitem{Mahm_C_Herman}
Mahmassani, H.S., G.-L.~Chang, and R.~Herman.  1986.  Individual Decisions
and Collective Effects in a Simulated Traffic System. \jrn{Transp.\ Sc.}
\rf 20(4) 86 258-271.

\bibitem{Axelrod}
Axelrod, R.  1984.  \jrn{The Evolution of Cooperation}.  New York: Basic
Books.

\bibitem{Hall}
Hall, F.L., B.L.~Allen, and M.A.~Gunter.  1986.  Empirical analysis of
freeway frow-density relationsships. \jrn{Transpn.\ Res.\ A} \rf 20A(3)
86 197--210.

\bibitem{k.axhausen}
Axhausen, K.  1994.  Personal communication.

\bibitem{varian}
MacKie-Mason, J.K., and H.R.~Varian, Pricing the internet, preprint 1994.

\bibitem{SOC}
Bak, P., C.~Tang, and K.~Wiesenfeld.  1988.  Self-organized criticality.
\jrn{Phys.\ Rev.\ A\/} \rf 38 88 368.

\bibitem{Duck}
Donald Duck, 1959, and related $\ldots$

\end{thebibliography}
\end{document}